# Visualizing quantum phenomena at complex oxide interfaces: an atomic view from scanning transmission electron microscopy

Hangwen Guo[1*], Mohammad Saghayezhian[2], Zhen Wang[2,3], Yimei Zhu[3], Jiandi Zhang[2], Ward Plummer[2*]

[1]*Institute for Nanoelectronic Devices and Quantum Computing, Fudan University, Shanghai 200433, China*

[2]*Department of Physics & Astronomy, Louisiana State University, Baton Rouge, LA 70803*

[3]*Condensed Matter Physics & Materials Science Department, Brookhaven National Laboratory, Upton, NY 11973*

**Complex oxide interfaces have been one of the central focuses in condensed matter physics and material science. Over the past decade, aberration corrected scanning transmission electron microscopy and spectroscopy has proven to be invaluable to visualize and understand the emerging quantum phenomena at an interface. In this paper, we briefly review some recent progress in the utilization of electron microscopy to probe interfaces. Specifically, we discuss several important challenges for electron microscopy to advance our understanding on interface phenomena, from the perspective of variable temperature, magnetism, electron energy loss spectroscopy analysis, electronic symmetry, and defects probing.**

**Keywords:** Complex oxide interfaces, scanning transmission electron microscopy, electron energy loss spectroscopy

*Correspondence should be addressed to: hangwenguo@fudan.edu.cn, wplummer@phys.lsu.edu

**Contents**

# 1 Introduction

Over the past decades, complex oxides have been one of the central and most exciting topics in the research of condensed matter physics. As compared to conventional semiconductors which are at the heart of contemporary industry, complex oxides exhibits richer physics and functionality. The origin of the diverse functionality is the entangled charge, spin, lattice and orbital degrees of freedom which have enabled many fascinating phenomena.

Complex oxide interfaces, on the other hand, are new arena to explore the novel functionalities [1–6]. Other than the aforementioned degrees of freedom, a key ingredient is the inversion symmetry breaking at an oxide interface (Fig. 1). The effect of the symmetry breaking is at least two-folded. On one hand, it triggers new interactions, for example, exchange bias effects by interfacing two magnetic layers [7–9]; two-dimensional electron gases by interfacing two insulators [10–13]; lattice or orbital reconstructions by interfacing two materials with different crystalline symmetries [14–17] etc. More interestingly, the symmetry breaking also facilitates the "cross-talking" among various degrees of freedom. A famous example is electric-field control of magnetism across magneto-electric interfaces [18–23]. Quoting the 2000 Nobel Prize winner Dr. Herber Kroemer: "Interface is the device" [24].

Understanding and controlling the rich phenomena at complex oxide interfaces not only demands the ability to fabricate high quality heterointerfaces [25–29], but also present great challenges for characterization. While comprehensive set of tools are often necessary to characterize and understand interfacial properties as shown in Fig. 1, the development of cross-sectional aberration corrected scanning transmission electron microscopy (STEM) has proven to be essential in studying interfaces [30]. The ability to identify site-by-site information with atomic resolution allows us to visualize the evolution of atomic displacements across the interfaces. Beyond precise determination of atomic position, the development of elemental core-level electron energy-loss spectroscopy (EELS) techniques provide local (column-by-column) chemical and electronic information, fundamentally enriching our knowledge of complex interfaces [31,32]. In this review, we present recent insights and challenges in understanding complex oxide interfaces, mostly in 3d perovskite oxides, by employing the STEM and EELS techniques.

The objective of this review is not to provide a complete and exhaustive overview of the existing literature on the oxide interface. Instead, we will focus on the following two major perspectives: (1) What advances have been made to identify the physical and chemical properties at complex oxide interfaces using electron microscopy and spectroscopy; (2) What are the challenges and issues in understanding the complex interfaces. We highlight several experimental results that shed light on these two perspectives. In particular, our main focus is to show the indispensability of STEM for understanding the properties of complex oxide interfaces and what needs to be improved and developed in the future. We note other techniques including synchrotron-based x-ray spectroscopies [33–37] (such as x-ray absorptions, x-ray linear dichroism and x-ray magnetic circular dichroism) and nonlinear optics [38–40] (such as second harmonic generations) are also powerful tools to study complex oxide interfaces. Combined with STEM-EELS techniques, one can often gather deeper understanding on the interface phenomena and mechanisms. We invite readers who are interested to consult the existing reviews and books [41,42].

This review is organized as follow: In section II, we briefly introduce the atomically resolved STEM techniques and discuss their application in oxide interfaces from three perspectives: structural distortions, chemical compositions, and valence states. In section III, we discuss several challenges and issues that we are facing in the community to understand oxide interfaces, from the perspective of variable temperature, magnetism, electronic symmetry, defects, and EELS analysis. In section IV, we provide a brief summary.

## 2 Advances in understanding complex oxide interfaces

2.1 A brief introduction of STEM

The combination of atomically-resolved STEM and EELS serve as powerful tools in determining the atomic and electronic structures in unprecedented details [43–45]. With the advent of spherical aberration correction and the usage of high brightness electron sources and post-specimen electron optics, the spatial resolution of STEM (electron probe) is near 50 picometer, which means both atomic structure and compositional/electronic structure are attainable atom by atom. By collecting electrons scattered into different angles, the microscope can resolve heavy elements, light element, as well as the strain information under different imaging mode. Since high-angle annular dark-field (HAADF) image collects electron scattered to large angles (Rutherford scattering), the atomic column intensity in an HAADF image is roughly proportional to $Z^2$ where Z is the atomic number, and hence it allows us to identify the atomic sites of heavy elements. Using an annular bright field (ABF) detector, the information of light element, such as oxygen ions, can be captured in the ABF-STEM images [30]. This is especially important since the oxygen displacements in complex oxides play crucial rule in their physical properties. The simultaneous collection of HAADF- and ABF- STEM images represents the sample structure over large defocus range and sample thicknesses. Low-angle ADF (LAADF) imaging emphasize scattering from lattice distortions, which is useful to visualize strain fields at interfaces or around defects [46,47].

The incident beam generates a number of signals while interaction with the sample, such as inelastic electrons, characteristic X-rays, secondary electrons (SE), and convergent beam electron diffraction (CBED) patterns *etc*. These signals can be collected simultaneously with the ADF-STEM images (or can be collected as a function of probe position), which provide information at atomic scale. Atomic-resolved STEM-SE imaging enables the selective visualization of surface features/atoms and bulk ones [48,49]. STEM-CBED, i.e. 4-dimensional (4D)-STEM imaging or DPC-STEM, has been utilized to map the internal electric field and magnetic structures from nanoscale to individual atoms [50–52]. STEM-Energy dispersive X-ray spectroscopy (EDX) and EELS in STEM mode allow us to map the elemental distribution at atomic-scale [53]. Beyond the elemental mapping, STEM-EELS also provide the bonding information and electronic structure at atomic-scale. In this review, we mainly focus on the STEM-EELS technique, which is a powerful tool to study oxide interfaces in recently years. The EELS spectrum is consisted of two regions, low-loss or valence (<50 eV) and high energy loss region (> 50eV) where the latter forms characteristic features that are element specific and equivalent to an absorption edge in X-ray absorption spectroscopy (XAS) [30–32].

2.2 STEM characterization on complex oxide interfaces

*2.2.1 Structural distortions*

In complex oxides, structural distortions usually hold the key to understand their electronic and magnetic properties [54,55]. For example, in *3d* transition metal oxides (TMOs), ferroelectric materials often have a polar distortion where metal and oxygen ions displace against one another. Such a structure is favored to have zero electrons in *3d* orbitals in TMOs, making them good insulators; on the other hand, magnetic and metallic *3d* TMOs often have antiferrodistortive (AFD) tilt distortions where oxygen octahedrons rotate [56,57]. In these materials, partially filled *d* orbitals lead to many exotic magnetic states and transport properties [57–59]. At high quality oxide interfaces, the requirement of connecting oxygen octahedrons will lead to modulations of their structural distortions [60]. Consequently, the physical properties can be altered. As discussed in section 2.1, atomically resolved STEM imaging allows us to identify the structural evolutions in a unit cell by unit cell manner, thus providing a unique opportunity to visualize oxide interfaces.

A recent work on ferroelectric $BaTiO_3$(BTO)/ferromagnetic $La_{0.67}Sr_{0.33}MnO_3$(LSMO) superlattices illustrates how STEM imaging can identify the structural evolutions across the interfaces and provide the information for the correlation between the structure and the physical properties [61]. In bulk form, BTO has a polar structure and LSMO has an AFD tilt structure. As shown in Fig. 2, when the thickness of both BTO and LSMO are relatively thick, both layers have their original structures. However, when the thickness of LSMO is reduced to 4 unit cell (u.c.), the structure of LSMO has completely changed to polar distortions. Further measurements reveal that such polar LSMO has stable ferromagnetism, thus serving as a potential candidate of rare multiferroics with ferromagnetic order [62]. Similar philosophy of "structural engineering via oxide interfaces" has yield many groundbreaking phenomena in recent years, such as: tuning magnetic anisotropy in LSMO and $SrRuO_3$ thin layers via interface-engineered octahedral environment [63,64]; controlling magnetism via symmetry mismatch [65]; manipulating electronic phase separations via chemical ordering [66]; room-temperature magnetoelectric multiferroic via lattice distortion engineering in $LuFe_2O_4$/$LuFeO_3$ superlattices [67]; artificial two-dimensional polar metals in $BaTiO_3$/$SrTiO_3$/$LaTiO_3$ superlattices [68]; perpendicular magnetic anisotropy induced by octahedral/tetrahedral modulations at LSMO/$LaCoO_{2.5}$ interfaces [14] etc. In some of these works, the bracketing materials belong to the same crystal structure class (such as perovskites) with different distorting octahedral patterns. In other works, the bracketing materials belong to distinct crystal structure class such as perovskite/brownmillerite [14], therefore the structural mismatch is more dramatic at interface. Although these results cover different aspects of physical properties in condensed matter physics, they all share a core feature that the structural distortions serve as the fundamental ingredient behind these phenomena. Without a doubt, atomic resolved STEM images provide the key information to support the conclusions.

*2.2.2 Chemical compositions*

Besides structural distortions, another important issue at oxide interfaces is the chemical compositions (intermixing). When two materials with different chemical elements are stacked together, a natural question is whether they are intermixed at interfaces and if so, to what extent? Fundamentally, such stoichiometry variations and intermixing are unavoidable in all complex oxide interfaces. This is due to the fact that the deposition of complex oxide films requires elevated temperatures, usually between 600~1000 °C [69,70]. In this temperature range, different elements are highly diffusive and mobile due to their thermodynamic nature so that the intermixing is not avoidable [29,71].

STEM-EELS is one of the best (if not the only) state-of-the-art technique to measure the interfacial ion intermixing with atomic resolution. Depending on the nature of interface and the optimization of growth condition, the level of intermixing can be controlled and minimized. At electronic polar and nonpolar interfaces such as $LaAlO_3/SrTiO_3$, Nakagawa *et al.* pointed out that the atomic disordering and stoichiometry changes can be significant due to the electronic nature of interfaces [72]. Similar observations have been reported for LSMO/STO interface, with tuning Sr doping level [73]. Meanwhile, in many systems with polar discontinuity, atomically sharp interface can be obtained, i.e. the stoichiometry intermixing is confined within one unit cell at the interface. Examples includes LSMO (with x=0.3)/BTO [61], LMO/LNO [74] etc. In most cases, the formation of intermixing is often asymmetric and complicated. It has been widely observed in many oxide systems that the level of cation and anion inter-diffusion depends on the stacking sequence of the oxide materials, with one interface sharper than the other [75–78]. The exact nature of such asymmetry is quite complex and beyond the scope of this review.

Most importantly, the power of STEM-EELS measurement is not only to identify the intermixing, but to provide the essential "missing piece" to understand the physical phenomena and mechanisms. An example is the influence of $LaAlO_3$ stoichiometry on the 2D electron gas formation at LAO/STO interfaces [79]. The emergent properties such as high mobility two-dimensional electron gas, magnetism and superconductivity between two insulators inspired numerous research interests [12,13,16,80–82]. However, the origin of such phenomena is still under debate. Warusawithana *et al.* placed their focus on the role of LAO stoichiometry [79], which was assumed to be nominal in most previous studies. They found out that only Al-rich LAO (with La/Al ratio about 0.9) will result in 2D conducting interface, while other stoichiometry (such as slightly La-rich) will result in insulating interface. Such stoichiometric variation is directly visualized with the aid of STEM-EELS measurements as shown in Fig. 3. This new observation provides a key perspective on understanding the nature of LAO/STO interface, which would be missing without STEM-EELS technique.

Another interesting example of the use of STEM is to tackle the "dead layer" problem in ultrathin manganite films. In many magnetic and conductive TMOs, at ultrathin limit, the TMOs are known to exhibit insulating and nonmagnetic behavior, losing their functional performance [7,83–87]. The origin of such "dead layer" effect has been under hot debates [88,89]. A typical example is LSMO with x = 0.3 where the metallic and ferromagnetic behavior disappear below critical thickness of 5 - 6 unit cell and it is believed that the stoichiometric variations are one of the key driving factors. Recently, Chen et al. investigated the microscopic origin of the stoichiometry by the combination of STEM-EELS, x-ray photoemission spectroscopy and other ex-situ measurement [90]. The STEM-EELS accurately determine the interfacial composition of different film thickness. The key finding is that below the film thickness of 10 u.c., the enhanced deviation of chemical composition is observed which is directly leading towards insulating transport behaviors.

It is worth noting that the STEM-EELS also shows its enormous power in the studies of systems other than TMOs. A noticeable example is the FeSe monolayer on STO surface which a superconducting transition temperature may be above liquid nitrogen temperature [91]. Zhao et al. discovered via HADDF-STEM and EELS measurements that an additional Se layer is also resolved to reside between the FeSe film and the $TiO_x$-terminated STO substrate [92]. Such extra Se layer along with the electron transfer plays the key role in enhancing the transition temperature, further demonstrating the unreplaceable importance of STEM-EELS technique at interfaces.

*2.2.3 Valence states*

Besides resolving stoichiometry and structural distortions, the ability to study the electronic valence information in a layer-by-layer fashion has tremendously advanced the understanding at oxide heterointerfaces. The multi-valence nature in transition metal ions has made them vulnerable to the surrounding environment. Several important phenomena have been observed by STEM-EELS regarding the electronic reconfiguration for the transition metal ions, including (but not limited to) electronic screening and charge transfer effects, etc. Many excellent works have been made in determining the valence states variation which supports the idea of screening and/or charge transfer effects [73,93,94]. For the scope of this review, in the following paragraph, we'd like to give an example where the mechanism of electrostatic carrier doping cannot be solely accounted for the observed EELS results, suggesting the necessity of deeper understanding the underlying physics at oxide interfaces as well as the interpretation of EELS data.

At magneto-electric interface, it is known that the different orientations of electric polarization from the ferroelectric material will result in different screening charges (i.e. electron or hole accumulations) at heterointerface, thus effectively tunes magnetic states at the magnetic oxides [19,63,95–97]. Such polarization induced local charge redistribution can be captured using STEM-EELS mapping technique. A noticeable example is the trilayer consisting of LSMO/PZT/LSMO [75]. Since the polarization of PZT is uniaxial and single domain, the bottom and top LSMO experiences opposite polarization fields at the interfaces as shown in Fig. 4. By measuring the EELS signal from *Mn-L$_{2,3}$* edge combined with *O-K* edge, the authors have mapped out the Mn valence evolution across both interfaces. As shown in Fig. 4c, at distances far from interface, the Mn valences have nominal values around +3.3 as in the bulk. Near the interface, both top and bottom LSMO layers exhibit reduction of the valence states with a larger drop about 0.26 over a deeper region at the top interface. These results are somewhat surprising and inconsistent with the well-known picture of screening effects. In this picture, the upward ferroelectric polarization would lead to electron accumulation in the top LSMO layer, and electron depletion in the bottom one. Therefore, one would expect that the Mn valence state will suppress at the top LSMO/PZT interface and increase at the bottom LSMO/PZT interface. Clearly, this is not the results from the EELS data in Fig. 4.

Such discrepancy may suggest the following: 1) Electrostatic doping only is often not enough to treat complex oxide interfaces. In complex oxides, strong Coulomb interaction and orbital hybridizations plays the key role at interface beyond the electrostatic picture [94]; 2) In most cases, any attempt at quantification of EELS experiments must be supported by theoretical simulations. The fine structure of EELS edges itself cannot be understood without access to the material's density of states (DOS) in the presence of a core hole, which can be provided by density functional theory (DFT) [53]. In section 3.3, we will elaborate these points with a few examples.

As a short summary for this section, we'd like to point out that aforementioned structural distortions, chemical composition and valence states are strongly entangled at oxide interfaces. For example, the compositional alternations at interfaces will lead to changes in metal-oxygen bond angles and bond length, thus further altering the orbital hybridizations and charge occupancies. Therefore, it is crucial that one needs to gain comprehensive knowledge by STEM-EELS characterization and other techniques to understand the phenomena.

# 3 Challenges in understanding complex oxide interfaces

## 3.1 Temperature dependent STEM

First, we'd like to mention that a variable-temperature atomic-resolved STEM and EELS is of critical importance for the complex oxide community, where temperature dependent structural transitions are common. Many phase transitions in superconducting, ferroelectric and magnetic materials often occur at low temperatures in complex oxides. The structure-property relationship near the transition temperature is still a major arena in correlated material community [98]. At oxide interface, an atomic-level low-temperature characterization technique is of urgent need. However, at low temperatures, many unfriendly factors become troublesome for STEM and EELS characterization including stage drift, sample stability and vibrations due to liquid nitrogen/helium bubbling/flow, which will blur the adapted images. Recent development of aberration-corrected cryo-STEM have shown a promising route towards this direction [99,100]. Atomically-resolved HAADF-STEM images of "charge-ordered" phase of bulk materials and STEM-EELS maps of LSMO/STO interfaces has been obtained at 100 K or liquid nitrogen temperature [101–103]. Cryo-STEM and EELS measurements across the FeSe/STO films were performed at 10 K and the results show that transferred electrons from STO accumulate within the first two layers of the FeSe film, which relates the significant enhanced $T_c$ of the film [92].

## 3.2 Resolving magnetic information

Interfaces have shown many emergent magnetic phenomena in recent years, ranging from interfacial ferromagnetism between two non-magnetic insulators to the interfacial magnetic skyrmions originated from Dzyaloshinskii-Moriya interactions and spin-orbit coupling [104–107]. While there are many methods to characterize the magnetic signals, the determination of the exact spin texture near interface regions is a challenging task. It is partially because the magnetic textures often contain both in-plane and out-of-plane components, therefore requiring nanoscale magnetic detections from different directions. In this section we briefly discuss the recent techniques and their pros and cons on this matter.

Lorentz TEM [108,109], electron holography [110,111], 4-D STEM / differential phase contrast (DPC)-STEM [112,113], and electron energy-loss magnetic chiral dichroism (EMCD) [112–114] represent the state-of-the-art techniques to detect magnetic textures in complex materials. Lorentz TEM has been employed directly to resolve the skyrmion spin texture in $Fe_{0.5}Co_{0.5}Si$ thin films [117] and studied magnetic transition from stripe to bubble state in $PbFe_{12}O_{19}$ [112]. Electron holography has identified the amplified magnetization near the antiphase boundary defects [111]. As shown in Fig. 5d and 5e, the combination of STEM imaging and DPC-STEM revealed one-to-one correspondence between local atomic structures and magnetic coupling of the twin boundaries in $Fe_3O_4$ [113]. Recently, the combination of EMCD and chromatic-aberration-corrected transmission electron microscopy is able to achieve atomic-level imaging of element-selective orbital and magnetic moments in $Sr_2FeMoO_6$ bulk system (Fig. 5a-d), which potentiate the study of magnetic interactions at atomic level [116].

For oxide interfaces, however, the above techniques are more challenging to detect interfacial magnetism as compared to bulk or thin film materials. One of the biggest issues to map out interfacial magnetic configuration is due to the STEM sample preparation. In order to detect the in-plane magnetic signal at interface, the electron beams need to be perpendicular to the surface/interface, therefore the sample must be grinded into very thin pieces (typically less than 20 nm) for the electron beam to pass through. So advanced grinding techniques and careful protections to prevent sample curling or bending are extremely important. The synthesis of free-standing perovskite films and heterostructures may provide a way to get thin and flat in-plane TEM samples in large size [118].

For out-of-plane magnetic signal, on the other hand, the major issue is the inherent electric potential perpendicular to the interface due to the symmetry breaking. Such electrostatic potential will be dominant at interface for both the over- and under-focus mode in Lorentz TEM, making it nearly impossible to resolve magnetic signal. Electron holography might be a possible choice, however, it will largely depend on how much electron wave phase shifts the spin texture at the interface. Moreover, atomic resolution has not been achieved yet. The newly developed 4D-STEM technique and EMCD with chromatic-aberration-corrected STEM may be a solution to resolve column-by-column magnetic textures [116].

3.3 EELS, Hybridization changes and charge transfer

The near-edge fine structures of core edges in EELS can be used to study the electronic structure of the materials. Therefore, it provides the opportunity to investigate the electronic structure in the local areas that are not accessible to any other experimental technique, such as grain boundaries, interfaces and extended dislocations. In the core-loss EELS, an incident electron inelastically interacts with a core electron, exciting the core electron to unoccupied states above Fermi energy, or to the continuum. The energy loss electron can be recorded by the spectrometer and provides information equivalent to absorption edge in X-ray absorption spectroscopy (XAS) [30–32]. The binding energy of core-loss EELS allow us to identify elements. The fine structure of the spectra is sensitive to local electronic states as well as coordination environment. This transition will result in a loss spectrum which is sensitive to local electronic states as well as coordination environment. The transition is mainly constrained by sum rules where the initial and final state should satisfy $\Delta l = \pm 1$, where l is orbital angular momentum. Therefore, depending on available density of states, Coulomb interaction between the excited electron and core hole and electron correlations, the intensity and energy of the EELS spectra varies. Proper interpretation of EELS data, beyond empirical trends, can elevate our understanding of the underlying nature of physical properties.

In EELS, in principle, it is possible to gain information about the nominal oxidation state of neighboring elements. However, the definition of nominal oxidation state has some degree of arbitrariness and often using elementary chemistry, assumes all bonds are perfectly ionic. Although, for example in oxides, nominal oxidation is meant to reflect the hybridization between the orbitals of the metal and oxygen, it does not provide direct information about the details of charge states on the respective atoms. The hybridization can lead to charge transfer from one atomic site to another, or it can effectively renormalize the energy states of two orbitals which does not necessitate any charge transfer. Therefore, change in the oxidation state does not mean transfer of electrons from one site to another [119]. A dramatic example are the calculations presented in references 54 and 126 showing no physical charge or charge transfer at interfaces of transition metal oxides.

It should be noted that EELS does not directly measure the nominal oxidation state, rather by following a systematic trend in metal and oxygen core edges sets boundaries for highest and lowest nominal oxygen states. For example, a systematic study of doped manganites, $La_{1-x}Ca_xMnO_3$ (LCMO) shows that with the increment of hole-doping, three features in EELS spectra systematically change, the Mn *$L_3/L_2$* edge ratio, normalized O K pre-peak and energy separation between the pre-peak and main peak of O K-edge [120]. Fig. 6a shows the Mn L edges for LCMO where a change in the intensity of *$L_2$* edge is seen as function of doping. Figure 6b shows the schematic for extraction of *$L_3$* and *$L_2$* intensities and their ratio is plotted in figure 6c against nominal oxidation state where a seemingly linear relation is observed. In Fig. 6d-f, a similar analysis is presented on *O-K* edge, where the intensity and relative peak positions of O K edge pre-peak and main peak are plotted against nominal oxidation state, which again a linear relation is seen [120]. While these analysis methods are not a measure of physical charge transfer, i.e. transfer of an electron going from $Mn^{3+}$ to $Mn^{4+}$, they are highly sensitive to orbital occupancy which is directly related to the degree of orbital hybridization between metal and oxygen [121]. The fact that the structure of LCMO remains orthorhombic with the same octahedral tilt and rotation (although different in magnitude) in all doping levels [122–125], removes the effect of structure on orbital hybridization in this material. Therefore, the only contributing factor will be the hole doping which can explain why the EELS data follow a linear trend. More studies are needed which aim to isolate the effect of structure and composition on orbital hybridization change. Thin films grown on different substrates with different symmetries can provide platforms in which while the composition is fixed, only the structure is modified.

While change in composition might change the charge state of the element, it is shown that nominal oxidation state does not need to follow the composition. There are instances where looking at the change in the measured composition directly shows that the charge counting does not correlate with nominal oxidation state and the idea of physical charge transfer is not accurate. For example, in the case of La$_{2/3}$Sr$_{1/3}$MnO$_3$/SrTiO$_3$ (LSMO/STO) (001), it has been shown that the concentration of La and Sr does not correlate with the nominal oxidation state derived from $L_{23}$ ratio [126]. Fig. 7a and 7b shows the HAADF and ABF image of LSMO/STO (001) where an abrupt interface is created. Although the corner shared octahedral are connected across the interface, due to the minor inevitable interface intermixture, the thin film finds its stoichiometric composition only after the first two unit cells. This short gradient in stoichiometry which is shown in Fig. 7c allows us to examine the validity of finding nominal oxidation state through charge counting. In this scheme, La$^{3+}$, Sr$^{2+}$ and O$^{2-}$ are the assumed charge states of the elements. Through charge neutrality condition, in La$_{2/3}$Sr$_{1/3}$MnO$_3$, Mn has must be in 3.3+ state. The composition of the first unit cell of LSMO in Fig. 7c is La$_{0.475}$Sr$_{0.525}$MnO$_3$ which requires the Mn to be in 3.5+ state, higher than the bulk value. However, figure 6d shows that the $L_{23}$ ratio near the interface shows an opposite trend where the nominal oxidation state is near 3+ and away from the interface becomes 3.3+ where the material recovers its stoichiometry. This discrepancy indicates that EELS is probing available density of states above Fermi energy (orbital occupancy) with a core hole present, and not the physical charge on each element. Therefore, care must be taken in interpreting the change in EELS core edges and these interpretations should be in conjunction with composition and structure change. More specifically, presence of composition intermixture, inevitable amount of oxygen vacancies, spin degree of freedom and core-hole effects (especially in insulators) present magnificent challenges for STEM-EELS analysis.

3.4 Electronic and structural symmetry at the interface

The ability to image the oxide interfaces in various projected directions is one of the triumphs of STEM. Broken symmetry at interfaces is exploited routinely to foster new physical properties at the junctions of two materials. To name a few, structural $\delta$- doping [127], induced in-plane anisotropy [63] and induced non-centrosymmetricity [61] have been quantified using ABF-STEM imaging. While valuable information has been obtained using this technique, assessing the in-plane symmetry of the atoms at the interface, which lies at the heart of a broken symmetry problem, remains a challenge. As an example, at the LSMO/STO (001) interface, two materials with different rotation patterns meet, as shown in Fig. 7. LSMO with $a^-a^-a^-$ symmetry and STO with $a^0a^0a^0$ symmetry might lead to considerable symmetry lowering at the interface. As shown in Fig. 8a and 8b, it has been determined from STEM data that only $a^-a^-c^0$ (c is perpendicular to the interface) rotation pattern can be responsible for the observed zig-zag oxygen arrangement [126,128]. Second harmonic generation is a technique that probes the electron symmetry at a region of broken symmetry [40]. Recent rotational SHG experiments have clearly demonstrated that the electron symmetry at the LSMO/STO interface is at best $C_2$ [129]. Using the STEM data shown in Fig. 7 and the SHG data, one can suggest an atomic arrangement at the interface consistent with STEM and SHG. Such a structure is shown for the $TiO_2$ plane in Fig. 8. Under $a^-a^-c^0$ rotation pattern, oxygen atoms move in and out of interface plane in $[001]_c$ (in pseudo-cubic notation). It is important to note that this type of atomic displacement is not directly attainable from the STEM images and the figure is created assuming that the octahedral rotation pattern remains unchanged, moving from the bulk of thin film to interface. Given the importance of interface-induced physical properties and its inherent broken symmetry, acquiring the exact atomic displacement as well as electronic symmetry will be of immense importance in determining the underlying physics in the observed phenomena. Combining STEM with SHG is one solution.

3.5 Defects at complex oxide interfaces

Last but not the least, a paramount challenge in understanding the complex oxide interface is to identify the impact of point defects on interfacial properties. As a matter of fact, this challenge not only exist at interfaces, but inside the thin film region or bulk crystal [130–132]. In solid state physics, defects can be categorized by its dimensionality, i.e. point defects and extended defects. Extended defects often have higher dimensionality (line, surface and volumetric) and exhibit non-equilibrium characteristics [133,134]. In many cases, STEM serves as a powerful technique to resolve those defects and their unique behaviors. Zurbuchen et al. have reviewed the morphology, structure and the nucleation process of two-dimensional out-of-phase boundaries (OPB) in epitaxial layered oxide films [135]. They point out that while the OPBs are rather random residing in the film, they do have tremendous impacts to decide film's functional properties. Random antiphase boundaries are known to be responsible for the anomalous magnetoresistive behavior in $Fe_3O_4$ and magnetic Heusler alloys, serving as critical factor for spintronics and memory device applications [136,137]. Our recent work has demonstrated an approach to create well-defined APB by growing perovskite oxide $La_{2/3}Sr_{1/3}MnO_3$ (LSMO113) thin film on unconventional Ruddlesden-Popper/$K_2NiF_4$-type $Sr_2RuO_4$ (SRO214) substrate [138]. This system holds a unique advantage since the substrate and film have a nearly perfect in-plane lattice match, essentially eliminating the constraint of substrate strain energy. Instead, the natural difference in stacking sequence between SRO214 and LSMO113 enables the nucleation of APB on the step at substrate/film interface to be resolved by STEM-EELS. Another work by Jeong et al. reported a rarely observed line defects formation in $NdTiO_3$ perovskite thin films by atomic-resolved STEM [139].

Point defects, on the other hand, are mainly governed by the second law of thermodynamics, therefore inevitably reside randomly in bulk and/or at interfaces [140,141]. Due to their zero-dimensional nature, it becomes extremely challenging to detect their existence when their densities are low. In complex oxides, one of the most common sources of point defects are oxygen vacancies. The effect of oxygen vacancies often play important roles at oxide interfaces due to the screening effects caused by non-canceling electric fields [93]. In contrast with monolayer TEM samples of 2-dimensional materials, where single atom vacancy can be directly observed (or clearly visualized) by STEM imaging, the TEM samples of oxide films are prepared to have certain thickness (usually thicker than 10 nm). Therefore, it is challenging to identify exact location and densities of the randomly distributed vacancies.

In recent years, oxygen vacancy can be detected and quantified via local lattice expansion [142] and slight change of EELS spectra of O-K edge [143,144], but with a detection limit of 1% from the latter approach [144]. We note that monochromated EELS with high energy resolution may be a possible choice to overcome the above issue. Recently, monochromated EELS in STEM with resolution below 20 meV has been achieved [145], which can locate and identify the point defects and its associated bandgap with a resolution of about 10nm in BAlGaN semiconductor [146]. In complex oxides, the high resolution monochromated EELS is also particularly important as it can detect subtle changes in the oxygen electronic configurations [147,148]. Combined with simulation, it could be a possible way to better "catch" the local oxygen vacancies, the associated phonon modes and bandgaps.

## 4 Summary

In this review, we have discussed how the atomically resolved STEM and EELS techniques have greatly benefited us to visualize and understand the quantum phenomena at complex oxide interfaces, from the perspective of structural distortions, compositions and electronic states. We also discuss the challenges and issues at understanding oxide interfaces. To tackle these problems, we suggest that comprehensive strategies are necessary by combining better spatial resolution, new detection methods and theoretical efforts. It is our firm belief that design and engineering of complex oxide interfaces will be brought to a new level in the near future.

**Acknowledgement**

This work was supported by the US Department of Energy (DOE) under Grant No. DOE DE-SC0002136. Z.W. and Y.Z. acknowledge support by the U.S. Department of Energy, Office of Basic Energy Science, Division of Materials Science and Engineering, under contract no. DESC0012704. H.G. acknowledge support by Shanghai Municipal Natural Science Foundation (19ZR1402800) and Shanghai Municipal Natural Science Foundation (18JC1411400).

**Competing interests**

The authors declare no competing interests.

**References**

[1] H. Y. Hwang, Y. Iwasa, M. Kawasaki, B. Keimer, N. Nagaosa, and Y. Tokura, Emergent phenomena at oxide interfaces, *Nature Materials* **11**, 103 (2012).

[2] J. Chakhalian, A. J. Millis, and J. Rondinelli, Whither the oxide interface, *Nature Materials* **11**, 92 (2012).

[3] J. Mannhart and D. G. Schlom, Oxide interfaces -- an opportunity for electronics, *Science* **327**, 1607 (2010).

[4] M. Bibes, E. Villegas Javier, and A. Barthélémy, Ultrathin oxide films and interfaces for electronics and spintronics, *Advances in Physics* **60**, 5 (2011).

[5] P. Zubko, S. Gariglio, M. Gabay, P. Ghosez, and J.-M. Triscone, Interface physics in complex oxide heterostructures, *Annual Review of Condensed Matter Physics* **2**, 141 (2011).

[6] M. Coll, J. Fontcuberta, M. Althammer, M. Bibes, H. Boschker, A. Calleja, G. Cheng, M. Cuoco, R. Dittmann, B. Dkhil, I. El Baggari, M. Fanciulli, I. Fina, E. Fortunato, C. Frontera, S. Fujita, V. Garcia, S. T. B. Goennenwein, C.-G. Granqvist, J. Grollier, R. Gross, A. Hagfeldt, G. Herranz, K. Hono, E. Houwman, M. Huijben, A. Kalaboukhov, D. J. Keeble, G. Koster, L. F. Kourkoutis, J. Levy, M. Lira-Cantu, J. L. MacManus-Driscoll, J. Mannhart, R. Martins, S. Menzel, T. Mikolajick, M. Napari, M. D. Nguyen, G. Niklasson, C. Paillard, S. Panigrahi, G. Rijnders, F. Sánchez, P. Sanchis, S. Sanna, D. G. Schlom, U. Schroeder, K. M. Shen, A. Siemon, M. Spreitzer, H. Sukegawa, R. Tamayo, J. van den Brink, N. Pryds, and F. M. Granozio, Towards oxide electronics: a roadmap, *Applied Surface Science* **482**, 1 (2019).

[7] M. Huijben, P. Yu, L. W. Martin, H. J. A. Molegraaf, Y.-H. Chu, M. B. Holcomb, N. Balke, G. Rijnders, and R. Ramesh, Ultrathin limit of exchange bias coupling at oxide multiferroic/ferromagnetic interfaces, *Advanced Materials* **25**, 4739 (2013).

[8] S. M. Wu, S. A. Cybart, P. Yu, M. D. Rossell, J. X. Zhang, R. Ramesh, and R. C. Dynes, Reversible electric control of exchange bias in a multiferroic field-effect device, *Nature Materials* **9**, 756 (2010).

[9] M. Gibert, P. Zubko, R. Scherwitzl, J. Íñiguez, and J.-M. Triscone, Exchange bias in LaNiO$_3$ –LaMnO$_3$ superlattices, *Nature Materials* **11**, 195 (2012).

[10] C. Cen, S. Thiel, J. Mannhart, and J. Levy, Oxide nanoelectronics on demand, *Science* **323**, 1026 (2009).

[11] A. Ohtomo and H. Y. Hwang, A high-mobility electron gas at the LaAlO$_3$/SrTiO$_3$ heterointerface, *Nature* **427**, 423 (2004).

[12] S. Thiel, G. Hammerl, A. Schmehl, C. W. Schneider, and J. Mannhart, Tunable quasi-two-dimensional electron gases in oxide heterostructures, *Science* **313**, 1942 (2006).

[13] N. Reyren, S. Thiel, A. D. Caviglia, L. F. Kourkoutis, G. Hammerl, C. Richter, C. W. Schneider, T. Kopp, A.-S. Rüetschi, D. Jaccard, M. Gabay, D. A. Muller, J.-M. Triscone, and J. Mannhart, Superconducting interfaces between insulating oxides, *Science* **317**, 1196 (2007).

[14] J. Zhang, Z. Zhong, X. Guan, X. Shen, J. Zhang, F. Han, H. Zhang, H. Zhang, X. Yan, Q. Zhang, L. Gu, F. Hu, R. Yu, B. Shen, and J. Sun, Symmetry mismatch-driven perpendicular magnetic anisotropy for perovskite/brownmillerite heterostructures, *Nature Communications* **9**, 1923 (2018).

[15] C. L. Jia, S. B. Mi, M. Faley, U. Poppe, J. Schubert, and K. Urban, Oxygen octahedron reconstruction in the SrTiO$_3$/LaAlO$_3$ heterointerfaces investigated using aberration-corrected ultrahigh-resolution transmission electron microscopy, *Physical Review B* **79**, 081405 (2009).

[16] M. Salluzzo, J. C. Cezar, N. B. Brookes, V. Bisogni, G. M. De Luca, C. Richter, S. Thiel, J. Mannhart, M. Huijben, A. Brinkman, G. Rijnders, and G. Ghiringhelli, Orbital Reconstruction and the Two-Dimensional Electron Gas at the LaAlO$_3$/SrTiO$_3$ interface, *Physical Review Letters* **102**, 166804 (2009).

[17] J. Chakhalian, J. W. Freeland, H.-U. Habermeier, G. Cristiani, G. Khaliullin, M. van Veenendaal, and B. Keimer, Orbital reconstruction and covalent bonding at an oxide interface, *Science* **318**, 1114 (2007).

[18] H. J. A. Molegraaf, J. Hoffman, C. A. F. Vaz, S. Gariglio, D. van der Marel, C.

H. Ahn, and J.-M. Triscone, Magnetoelectric effects in complex oxides with competing ground states, *Advanced Materials* **21**, 3470 (2009).

[19] C. A. F. Vaz, J. Hoffman, C. H. Ahn, and R. Ramesh, Magnetoelectric coupling effects in multiferroic complex oxide composite structures, *Advanced Materials* **22**, 2900 (2010).

[20] C. A. F. Vaz, Electric field control of magnetism in multiferroic heterostructures, *Journal of Physics: Condensed Matter* **24**, 333201 (2012).

[21] W. Eerenstein, N. D. Mathur, and J. F. Scott, Multiferroic and magnetoelectric materials, *Nature* **442**, 759 (2006).

[22] S. Dong, J.-M. Liu, S.-W. Cheong, and Z. Ren, Multiferroic materials and magnetoelectric physics: symmetry, entanglement, excitation, and topology, *Advances in Physics* **64**, 519 (2015).

[23] W. Huang, Y. Yin, and X. Li, Atomic-scale mapping of interface reconstructions in multiferroic heterostructures, *Applied Physics Reviews* **5**, 41110 (2018).

[24] H. Kroemer, Nobel Lecture: Quasielectric fields and band offsets: teaching electrons new tricks, *Reviews of Modern Physics* **73**, 783 (2001).

[25] D. P. Norton, Synthesis and properties of epitaxial electronic oxide thin-film materials, *Materials Science and Engineering: R: Reports* **43**, 139 (2004).

[26] P. R. Willmott, Deposition of complex multielemental thin films, *Progress in Surface Science* **76**, 163 (2004).

[27] M. N. R. Ashfold, F. Claeyssens, G. M. Fuge, and S. J. Henley, Pulsed laser ablation and deposition of thin films, *Chemical Society Reviews* **33**, 23 (2004).

[28] H. M. Christian and G. Eres, Recent advances in pulsed-laser deposition of complex oxides, *Journal of Physics: Condensed Matter* **20**, 264005 (2008).

[29] H. Guo, D. Sun, W. Wang, Z. Gai, I. Kravchenko, J. Shao, L. Jiang, T. Z. Ward, P. C. Snijders, L. Yin, J. Shen, and X. Xu, Growth diagram of $La_{0.7}Sr_{0.3}MnO_3$ thin films using pulsed laser deposition, *Journal of Applied Physics* **113**, 234301 (2013).

[30] M. Varela, A. R. Lupini, K. van Benthem, A. Y. Borisevich, M. F. Chisholm,

N. Shibata, E. Abe, and S. J. Pennycook, Materials characterization in the aberration-corrected scanning transmission electron microscope, *Annual Review of Materials Research* **35**, 539 (2005).

[31] R. F. Egerton, Electron energy-loss spectroscopy in the TEM, *Reports on Progress in Physics* **72**, 016502 (2009).

[32] F. Hofer, F. P. Schmidt, W. Grogger, and G. Kothleitner, Fundamentals of electron energy-loss spectroscopy, *IOP Conference Series: Materials Science and Engineering* **109**, 012007 (2016).

[33] Y. Cao, X. Liu, M. Kareev, D. Choudhury, S. Middey, D. Meyers, J.-W. Kim, P. J. Ryan, J. W. Freeland, and J. Chakhalian, Engineered Mott ground state in a $LaTiO_{3+\delta}$/$LaNiO_3$ heterostructure, *Nature Communications* **7**, 10418 (2016).

[34] X. Chi, Z. Huang, T. C. Asmara, K. Han, X. Yin, X. Yu, C. Diao, M. Yang, D. Schmidt, P. Yang, P. E. Trevisanutto, T. J. Whitcher, T. Venkatesan, M. B. H. Breese, Ariando, and A. Rusydi, Large enhancement of 2D electron gases mobility induced by interfacial localized electron screening effect, *Advanced Materials* **30**, 1707428 (2018).

[35] G. Fabbris, N. Jaouen, D. Meyers, J. Feng, J. D. Hoffman, R. Sutarto, S. G. Chiuzbăian, A. Bhattacharya, and M. P. M. Dean, Emergent c-axis magnetic helix in manganite-nickelate superlattices, *Physical Review B* **98**, 180401 (2018).

[36] F. Y. Bruno, M. N. Grisolia, C. Visani, S. Valencia, M. Varela, R. Abrudan, J. Tornos, A. Rivera-Calzada, A. A. Ünal, S. J. Pennycook, Z. Sefrioui, C. Leon, J. E. Villegas, J. Santamaria, A. Barthélémy, and M. Bibes, Insight into spin transport in oxide heterostructures from interface-resolved magnetic mapping, *Nature Communications* **6**, 6306 (2015).

[37] X. Zhai, L. Cheng, Y. Liu, C. M. Schlepütz, S. Dong, H. Li, X. Zhang, S. Chu, L. Zheng, J. Zhang, A. Zhao, H. Hong, A. Bhattacharya, J. N. Eckstein, and C. Zeng, Correlating interfacial octahedral rotations with magnetism in $(LaMnO_{3+\delta})_N/(SrTiO_3)_N$ superlattices, *Nature Communications* **5**, 4283 (2014).

[38] A. Rubano, G. De Luca, J. Schubert, Z. Wang, S. Zhu, D. G. Schlom, L.

Marrucci, and D. Paparo, Polar asymmetry of La$_{(1-\delta)}$Al$_{(1+\delta)}$O$_3$/SrTiO$_3$ heterostructures probed by optical second harmonic generation, *Applied Physics Letters* **107**, 101603 (2015).

[39] S. Middey, P. Rivero, D. Meyers, M. Kareev, X. Liu, Y. Cao, J. W. Freeland, S. Barraza-Lopez, and J. Chakhalian, Polarity compensation in ultra-thin films of complex oxides: The case of a perovskite nickelate, *Scientific Reports* **4**, 6819 (2014).

[40] D. Paparo, A. Rubano, and L. Marrucci, Optical second-harmonic generation selection rules and resonances in buried oxide interfaces: the case of LaAlO$_3$/SrTiO$_3$, *Journal of the Optical Society of America B* 30, 2452 (2013).

[41] Claudia Cancellieri and Vladimir Strocov, editors, *Spectroscopy of Complex Oxide Interfaces* (Springer International Publishing, 2018).

[42] F. Manfred, Phase engineering in oxides by interfaces, *Philosophical Transactions of the Royal Society A: Mathematical, Physical and Engineering Sciences* **370**, 4972 (2012).

[43] Stephen J. Pennycook and Peter D. Nellist, editors, *Scanning Transmission Electron Microscopy*, 1st ed. (Springer-Verlag New York, 2011).

[44] S. J. Pennycook, M. F. Chisholm, A. R. Lupini, M. Varela, A. Y. Borisevich, M. P. Oxley, W. D. Luo, K. van Benthem, S.-H. Oh, D. L. Sales, S. I. Molina, J. García-Barriocanal, C. Leon, J. Santamaría, S. N. Rashkeev, and S. T. Pantelides, Aberration-corrected scanning transmission electron microscopy: from atomic imaging and analysis to solving energy problems, *Philosophical Transactions of the Royal Society A: Mathematical, Physical and Engineering Sciences* **367**, 3709 (2009)

[45] S. J. Pennycook, Seeing the atoms more clearly: STEM imaging from the Crewe era to today, *Ultramicroscopy* **123**, 28 (2012).

[46] Z. Yu, D. A. Muller, and J. Silcox, Study of strain fields at a-Si/c-Si interface, *Journal of Applied Physics* **95**, 3362 (2004).

[47] P. J. Phillips, M. De Graef, L. Kovarik, A. Agrawal, W. Windl, and M. J. Mills, Atomic-resolution defect contrast in low angle annular dark-field STEM,

*Ultramicroscopy* **116**, 47 (2012).

[48] Y. Zhu, H. Inada, K. Nakamura, and J. Wall, Imaging single atoms using secondary electrons with an aberration-corrected electron microscope, *Nature Materials* **8**, 808 (2009).

[49] H. Inada, K. Tamura, K. Nakamura, Y. Suzuki, M. Konno, D. Su, J. Wall, R. Egerton, and Y. Zhu, Atomic resolved secondary electron imaging with an aberration corrected scanning transmission electron microscope, *Microscopy and Microanalysis* **17**, 1298 (2011).

[50] J. A. Hachtel, J. C. Idrobo, and M. Chi, Sub-Ångstrom electric field measurements on a universal detector in a scanning transmission electron microscope, *Advanced Structural and Chemical Imaging* **4**, 10 (2018).

[51] S. Das, Y. L. Tang, Z. Hong, M. A. P. Gonçalves, M. R. McCarter, C. Klewe, K. X. Nguyen, F. Gómez-Ortiz, P. Shafer, E. Arenholz, V. A. Stoica, S.-L. Hsu, B. Wang, C. Ophus, J. F. Liu, C. T. Nelson, S. Saremi, B. Prasad, A. B. Mei, D. G. Schlom, J. Íñiguez, P. García-Fernández, D. A. Muller, L. Q. Chen, J. Junquera, L. W. Martin, and R. Ramesh, Observation of room-temperature polar skyrmions, *Nature* **568**, 368 (2019).

[52] C. Ophus, Four-Dimensional Scanning Transmission Electron Microscopy (4D-STEM): From Scanning Nanodiffraction to Ptychography and Beyond, *Microscopy and Microanalysis* **25**, 563 (2019).

[53] P. Gao, R. Ishikawa, B. Feng, A. Kumamoto, N. Shibata, and Y. Ikuhara, Atomic-scale structure relaxation, chemistry and charge distribution of dislocation cores in $SrTiO_3$, *Ultramicroscopy* **184**, 217 (2018).

[54] M. B. Salamon and M. Jaime, The physics of manganites: Structure and transport, *Reviews of Modern Physics* **73**, 583 (2001).

[55] G. Koster, L. Klein, W. Siemons, G. Rijnders, J. S. Dodge, C.-B. Eom, D. H. A. Blank, and M. R. Beasley, Structure, physical properties, and applications of $SrRuO_3$ thin films, *Reviews of Modern Physics* **84**, 253 (2012).

[56] J. Kim, Y. Kim, Y. S. Kim, J. Lee, L. Kim, and D. Jung, Large nonlinear dielectric properties of artificial $BaTiO_3$/$SrTiO_3$ superlattices, *Applied Physics*

*Letters* **80**, 3581 (2002).

[57] M.-H. Whangbo, E. E. Gordon, J. L. Bettis, A. Bussmann-Holder, and J. Köhler, Tolerance factor and cation–anion orbital interactions differentiating the polar and antiferrodistortive structures of perovskite oxides ABO$_3$, *Zeitschrift Für Anorganische Und Allgemeine Chemie* **641**, 1043 (2015).

[58] W. Zhong and D. Vanderbilt, Competing structural istabilities in cubic perovskites, *Physical Review Letters* **74**, 2587 (1995).

[59] U. Aschauer and N. A. Spaldin, Competition and cooperation between antiferrodistortive and ferroelectric instabilities in the model perovskite SrTiO$_3$, *Journal of Physics: Condensed Matter* **26**, 122203 (2014).

[60] J. M. Rondinelli, S. J. May, and J. W. Freeland, Control of octahedral connectivity in perovskite oxide heterostructures: An emerging route to multifunctional materials discovery, *MRS Bulletin* **37**, 261 (2012).

[61] H. Guo, Z. Wang, S. Dong, S. Ghosh, M. Saghayezhian, L. Chen, Y. Weng, A. Herklotz, T. Z. Ward, R. Jin, S. T. Pantelides, Y. Zhu, J. Zhang, and E. W. Plummer, Interface-induced multiferroism by design in complex oxide superlattices, *Proceedings of the National Academy of Sciences of the United States of America* **114**, E5062 (2017).

[62] J. H. Lee, L. Fang, E. Vlahos, X. Ke, Y. W. Jung, L. F. Kourkoutis, J.-W. Kim, P. J. Ryan, T. Heeg, M. Roeckerath, V. Goian, M. Bernhagen, R. Uecker, P. C. Hammel, K. M. Rabe, S. Kamba, J. Schubert, J. W. Freeland, D. A. Muller, C. J. Fennie, P. Schiffer, V. Gopalan, E. Johnston-Halperin, and D. G. Schlom, A strong ferroelectric ferromagnet created by means of spin-lattice coupling, *Nature* **466**, 954 (2010).

[63] Z. Liao, M. Huijben, Z. Zhong, N. Gauquelin, S. Macke, R. J. Green, S. Van Aert, J. Verbeeck, G. Van Tendeloo, K. Held, G. A. Sawatzky, G. Koster, and G. Rijnders, Controlled lateral anisotropy in correlated manganite heterostructures by interface-engineered oxygen octahedral coupling, *Nature Materials* **15**, 425 (2016).

[64] D. Kan, R. Aso, R. Sato, M. Haruta, H. Kurata, and Y. Shimakawa, Tuning

magnetic anisotropy by interfacially engineering the oxygen coordination environment in a transition metal oxide, *Nature Materials* **15**, 432 (2016).

[65] E.-J. Guo, R. Desautels, D. Lee, M. A. Roldan, Z. Liao, T. Charlton, H. Ambaye, J. Molaison, R. Boehler, D. Keavney, A. Herklotz, T. Z. Ward, H. N. Lee, and M. R. Fitzsimmons, Exploiting symmetry mismatch to control magnetism in a ferroelastic heterostructure, *Physical Review Letters* **122**, 187202 (2019).

[66] Y. Zhu, K. Du, J. Niu, L. Lin, W. Wei, H. Liu, H. Lin, K. Zhang, T. Yang, Y. Kou, J. Shao, X. Gao, X. Xu, X. Wu, S. Dong, L. Yin, and J. Shen, Chemical ordering suppresses large-scale electronic phase separation in doped manganites, *Nature Communications* **7**, 11260 (2016).

[67] J. A. Mundy, C. M. Brooks, M. E. Holtz, J. A. Moyer, H. Das, A. F. Rébola, J. T. Heron, J. D. Clarkson, S. M. Disseler, Z. Liu, A. Farhan, R. Held, R. Hovden, E. Padgett, Q. Mao, H. Paik, R. Misra, L. F. Kourkoutis, E. Arenholz, A. Scholl, J. A. Borchers, W. D. Ratcliff, R. Ramesh, C. J. Fennie, P. Schiffer, D. A. Muller, and D. G. Schlom, Atomically engineered ferroic layers yield a room-temperature magnetoelectric multiferroic, *Nature* **537**, 523 (2016).

[68] Y. Cao, Z. Wang, S. Y. Park, Y. Yuan, X. Liu, S. M. Nikitin, H. Akamatsu, M. Kareev, S. Middey, D. Meyers, P. Thompson, P. J. Ryan, P. Shafer, A. N'Diaye, E. Arenholz, V. Gopalan, Y. Zhu, K. M. Rabe, and J. Chakhalian, Artificial two-dimensional polar metal at room temperature, *Nature Communications* **9**, 1547 (2018).

[69] G. Koster, M. Huijben, and G. Rijnders, editors , *Epitaxial Growth of Complex Metal Oxides*, 1st ed. (Elsevier, Woodhead Publishing Etd, 2015).

[70] D. B. Chrisey and G. K. Hubler, *Pulse Laser Deposition of Thin Films* (1994).

[71] I. V. Markov, *Crystal Growth for Beginners: Fundamentals of Nucleation, Crystal Growth and Epitaxy*, 2nd ed. (World Scientific Publishing Company, 2003).

[72] N. Nakagawa, H. Y. Hwang, and D. A. Muller, Why some interfaces cannot be sharp, *Nature Materials* **5**, 204 (2006).

[73] J. A. Mundy, Y. Hikita, T. Hidaka, T. Yajima, T. Higuchi, H. Y. Hwang, D. A. Muller, and L. F. Kourkoutis, Visualizing the interfacial evolution from charge compensation to metallic screening across the manganite metal–insulator transition, *Nature Communications* **5**, 3464 (2014).

[74] M. Gibert, M. Viret, A. Torres-Pardo, C. Piamonteze, P. Zubko, N. Jaouen, J.-M. Tonnerre, A. Mougin, J. Fowlie, S. Catalano, A. Gloter, O. Stéphan, and J.-M. Triscone, Interfacial control of magnetic properties at $LaMnO_3$/$LaNiO_3$ interfaces, *Nano Letters* **15**, 7355 (2015).

[75] S. R. Spurgeon, P. V Balachandran, D. M. Kepaptsoglou, A. R. Damodaran, J. Karthik, S. Nejati, L. Jones, H. Ambaye, V. Lauter, Q. M. Ramasse, K. K. S. Lau, L. W. Martin, J. M. Rondinelli, and M. L. Taheri, Polarization screening-induced magnetic phase gradients at complex oxide interfaces, *Nature Communications* **6**, 6735 (2015).

[76] J. H. Lee, G. Luo, I. C. Tung, S. H. Chang, Z. Luo, M. Malshe, M. Gadre, A. Bhattacharya, S. M. Nakhmanson, J. A. Eastman, H. Hong, J. Jellinek, D. Morgan, D. D. Fong, and J. W. Freeland, Dynamic layer rearrangement during growth of layered oxide films by molecular beam epitaxy, *Nature Materials* **13**, 879 (2014).

[77] M. Arredondo, M. Saunders, A. Petraru, H. Kohlstedt, I. Vrejoiu, M. Alexe, D. Hesse, N. D. Browning, P. Munroe, and V. Nagarajan, Structural defects and local chemistry across ferroelectric–electrode interfaces in epitaxial heterostructures, *Journal of Materials Science* **44**, 5297 (2009).

[78] L. F. Kourkoutis, D. A. Muller, Y. Hotta, and H. Y. Hwang, Asymmetric interface profiles in $LaVO_3$/$SrTiO_3$ heterostructures grown by pulsed laser deposition, *Applied Physics Letters* **91**, 163101 (2007).

[79] M. P. Warusawithana, C. Richter, J. A. Mundy, P. Roy, J. Ludwig, S. Paetel, T. Heeg, A. A. Pawlicki, L. F. Kourkoutis, M. Zheng, M. Lee, B. Mulcahy, W. Zander, Y. Zhu, J. Schubert, J. N. Eckstein, D. A. Muller, C. S. Hellberg, J. Mannhart, and D. G. Schlom, $LaAlO_3$ stoichiometry is key to electron liquid formation at $LaAlO_3$/$SrTiO_3$ interfaces, *Nature Communications* **4**, 2351

(2013).

[80] M. Sing, G. Berner, K. Goß, A. Müller, A. Ruff, A. Wetscherek, S. Thiel, J. Mannhart, S. A. Pauli, C. W. Schneider, P. R. Willmott, M. Gorgoi, F. Schäfers, and R. Claessen, Profiling the interface electron gas of LaAlO$_3$/SrTiO$_3$ heterostructures with hard x-tay photoelectron spectroscopy, *Physical Review Letters* **102**, 176805 (2009).

[81] J. A. Bert, B. Kalisky, C. Bell, M. Kim, Y. Hikita, H. Y. Hwang, and K. A. Moler, Direct imaging of the coexistence of ferromagnetism and superconductivity at the LaAlO$_3$/SrTiO$_3$ interface, *Nature Physics* **7**, 767 (2011).

[82] Z. Zhong, P. X. Xu, and P. J. Kelly, Polarity-induced oxygen vacancies at interfaces, *Physical Review B* **82**, 165127 (2010).

[83] P. Maksymovych, M. Huijben, M. Pan, S. Jesse, N. Balke, Y.-H. Chu, H. J. Chang, A. Y. Borisevich, A. P. Baddorf, G. Rijnders, D. H. A. Blank, R. Ramesh, and S. V Kalinin, Ultrathin limit and dead-layer effects in local polarization switching of BiFeO$_3$, *Physical Review B* **85**, 014119 (2012).

[84] J. Xia, W. Siemons, G. Koster, M. R. Beasley, and A. Kapitulnik, Critical thickness for itinerant ferromagnetism in ultrathin films of SrRuO$_3$, *Physical Review B* **79**, 140407 (2009).

[85] M. Huijben, L. W. Martin, Y.-H. Chu, M. B. Holcomb, P. Yu, G. Rijnders, D. H. A. Blank, and R. Ramesh, critical thickness and orbital ordering in ultrathin La$_{0.7}$Sr$_{0.3}$MnO$_3$ films, *Physical Review B* **78**, 094413 (2008).

[86] J. Junquera and P. Ghosez, Critical thickness for ferroelectricity in perovskite ultrathin films, *Nature* **422**, 506 (2003).

[87] Z. Liao, F. Li, P. Gao, L. Li, J. Guo, X. Pan, R. Jin, E. W. Plummer, and J. Zhang, Origin of the metal-insulator transition in ultrathin films of La$_{2/3}$Sr$_{1/3}$MnO$_3$, *Physical Review B* **92**, 125123 (2015).

[88] E.-J. Guo, M. A. Roldan, T. Charlton, Z. Liao, Q. Zheng, H. Ambaye, A. Herklotz, Z. Gai, T. Z. Ward, H. N. Lee, and M. R. Fitzsimmons, Removal of the magnetic dead layer by geometric design, *Advanced Functional Materials*

**28**, 1800922 (2018).

[89] G. Wang, Z. Wang, M. Meng, M. Saghayezhian, L. Chen, C. Chen, H. Guo, Y. Zhu, E. W. Plummer, and J. Zhang, Role of disorder and correlations in the metal-insulator transition in ultrathin $SrVO_3$ films, *Physical Review B* **100**, 155114 (2019).

[90] L. Chen, Z. Wang, G. Wang, H. Guo, M. Saghayezhian, J. Tao, Y. Zhu, E.W. Plummer, and J. Zhang, Surface and interface properties of $La_{2/3}Sr_{1/3}MnO_3$ thin films on $SrTiO_3$ (001), *Physical Review Materials* **3**, 044407 (2019).

[91] J.-F. Ge, Z.-L. Liu, C. Liu, C.-L. Gao, D. Qian, Q.-K. Xue, Y. Liu, and J.-F. Jia, Superconductivity above 100 K in single-layer FeSe films on doped $SrTiO_3$, *Nature Materials* **14**, 285 (2014).

[92] W. Zhao, M. Li, C.-Z. Chang, J. Jiang, L. Wu, C. Liu, J. S. Moodera, Y. Zhu, and M. H. W. Chan, Direct imaging of electron transfer and its influence on superconducting pairing at $FeSe/SrTiO_3$ interface, *Science Advances* **4**, eaao2682 (2018).

[93] Y.-M. Kim, A. Morozovska, E. Eliseev, M. P. Oxley, R. Mishra, S. M. Selbach, T. Grande, S. T. Pantelides, S. V Kalinin, and A. Y. Borisevich, Direct observation of ferroelectric field effect and vacancy-controlled screening at the $BiFeO_3/La_xSr_{1-x}MnO_3$ interface, *Nature Materials* **13**, 1019 (2014).

[94] M. N. Grisolia, J. Varignon, G. Sanchez-Santolino, A. Arora, S. Valencia, M. Varela, R. Abrudan, E. Weschke, E. Schierle, J. E. Rault, J.-P. Rueff, A. Barthelemy, J. Santamaria, and M. Bibes, Hybridization-controlled charge transfer and induced magnetism at correlated oxide interfaces, *Nature Physics* **12**, 484 (2016).

[95] M. Izumi, Y. Ogimoto, Y. Okimoto, T. Manako, P. Ahmet, K. Nakajima, T. Chikyow, M. Kawasaki, and Y. Tokura, Insulator-metal transition induced by interlayer coupling in $La_{0.6}Sr_{0.4}MnO_3/SrTiO_3$ superlattices, *Physical Review B* **64**, 064429 (2001).

[96] S. Dong, X. Zhang, R. Yu, J.-M. Liu, and E. Dagotto, Microscopic model for the ferroelectric field effect in oxide heterostructures, *Physical Review B* **84**,

155117 (2011).

[97] L. Jiang, W. S. Choi, H. Jeen, S. Dong, Y. Kim, M.-G. Han, Y. Zhu, S. V Kalinin, E. Dagotto, T. Egami, and H. N. Lee, Tunneling electroresistance induced by interfacial phase transitions in ultrathin oxide heterostructures, *Nano Letters* **13**, 5837 (2013).

[98] D. Lee, B. Chung, Y. Shi, G.-Y. Kim, N. Campbell, F. Xue, K. Song, S.-Y. Choi, J. P. Podkaminer, T. H. Kim, P. J. Ryan, J.-W. Kim, T. R. Paudel, J.-H. Kang, J. W. Spinuzzi, D. A. Tenne, E. Y. Tsymbal, M. S. Rzchowski, L. Q. Chen, J. Lee, and C. B. Eom, Isostructural metal-insulator transition in $VO_2$, *Science* **362**, 1037 (2018).

[99] L. F. Kourkoutis, I. El Baggari, B. H. Savitzky, D. J. Baek, B. H. Goodge, R. Hovden, and M. J. Zachman, Aberration-corrected STEM/EELS at cryogenic temperatures, *Microscopy and Microanalysis* **23**, 428 (2017).

[100] I. El Baggari, G. M. Stiehl, J. Waelder, D. C. Ralph, and L. F. Kourkoutis, Atomic-resolution Cryo-STEM Imaging of a Structural Phase Transition in $TaTe_2$, *Microscopy and Microanalysis* **24**, 86 (2018).

[101] I. El Baggari, D. J. Baek, B. H. Savitzky, M. J. Zachman, R. Hovden, and L. F. Kourkoutis, Low temperature electron microscopy of "charge-ordered" phases, *Microscopy and Microanalysis* **25**, 934 (2019).

[102] I. El Baggari, B. H. Savitzky, A. S. Admasu, J. Kim, S.-W. Cheong, R. Hovden, and L. F. Kourkoutis, Nature and evolution of incommensurate charge order in manganites visualized with cryogenic scanning transmission electron microscopy, *Proceedings of the National Academy of Sciences of the United States of America* **115**, 1445 (2018).

[103] B. H. Goodge, D. J. Baek, and L. F. Kourkoutis, Direct electron detection for atomic resolution in situ EELS, *Microscopy and Microanalysis* **24**, 1844 (2018).

[104] A. Brinkman, M. Huijben, M. van Zalk, J. Huijben, U. Zeitler, J. C. Maan, W. G. van der Wiel, G. Rijnders, D. H. A. Blank, and H. Hilgenkamp, Magnetic effects at the interface between non-magnetic oxides, *Nature Materials* **6**, 493

(2007).

[105] L. Wang, Q. Feng, Y. Kim, R. Kim, K. H. Lee, S. D. Pollard, Y. J. Shin, H. Zhou, W. Peng, D. Lee, W. Meng, H. Yang, J. H. Han, M. Kim, Q. Lu, and T. W. Noh, Ferroelectrically tunable magnetic skyrmions in ultrathin oxide heterostructures, *Nature Materials* **17**, 1087 (2018).

[106] J. Matsuno, N. Ogawa, K. Yasuda, F. Kagawa, Interface-driven topological Hall effect in $SrRuO_3$/$SrIrO_3$ bilayer, W. Koshibae, N. Nagaosa, Y. Tokura, and M. Kawasaki, *Science Advances* **2**, e1600304 (2016).

[107] F. Hellman, A. Hoffmann, Y. Tserkovnyak, G. S. D. Beach, E. E. Fullerton, C. Leighton, A. H. MacDonald, D. C. Ralph, D. A. Arena, H. A. Dürr, P. Fischer, J. Grollier, J. P. Heremans, T. Jungwirth, A. V Kimel, B. Koopmans, I. N. Krivorotov, S. J. May, A. K. Petford-Long, J. M. Rondinelli, N. Samarth, I. K. Schuller, A. N. Slavin, M. D. Stiles, O. Tchernyshyov, A. Thiaville, and B. L. Zink, Interface-induced phenomena in magnetism, *Reviews of Modern Physics* **89**, 25006 (2017).

[108] C. Phatak, A. K. Petford-Long, and M. De Graef, Recent advances in Lorentz microscopy, *Current Opinion in Solid State and Materials Science* **20**, 107 (2016).

[109] D.-T. Ngo and L. T. Kuhn, In situ transmission electron microscopy for magnetic nanostructures, *Advances in Natural Sciences: Nanoscience and Nanotechnology* **7**, 45001 (2016).

[110] A. Kovács and R. E. Dunin-Borkowski, *Magnetic Imaging of Nanostructures Using Off-Axis Electron Holography* (Elsevier, 2018, edited by E. Brück), pp. 59–153.

[111] Y. Murakami, K. Niitsu, T. Tanigaki, R. Kainuma, H. S. Park, and D. Shindo, Magnetization amplified by structural disorder within nanometre-scale interface region, *Nature Communications* **5**, 4133 (2014).

[112] J. Zweck, Imaging of magnetic and electric fields by electron microscopy, *Journal of Physics: Condensed Matter* **28**, 403001 (2016).

[113] C. Chen, H. Li, T. Seki, D. Yin, G. Sanchez-Santolino, K. Inoue, N. Shibata,

and Y. Ikuhara, Direct determination of atomic structure and magnetic coupling of magnetite twin boundaries, *ACS Nano* **12**, 2662 (2018).

[114] Z. Q. Wang, X. Y. Zhong, R. Yu, Z. Y. Cheng, and J. Zhu, Quantitative experimental determination of site-specific magnetic structures by transmitted electrons, *Nature Communications* **4**, 1395 (2013).

[115] P. Schattschneider, S. Rubino, C. Hébert, J. Rusz, J. Kuneš, P. Novák, E. Carlino, M. Fabrizioli, G. Panaccione, and G. Rossi, Detection of magnetic circular dichroism using a transmission electron microscope, *Nature* **441**, 486 (2006).

[116] Z. Wang, A. H. Tavabi, L. Jin, J. Rusz, D. Tyutyunnikov, H. Jiang, Y. Moritomo, J. Mayer, R. E. Dunin-Borkowski, R. Yu, J. Zhu, and X. Zhong, Atomic scale imaging of magnetic circular dichroism by achromatic electron microscopy, *Nature Materials* **17**, 221 (2018).

[117] X. Z. Yu, Y. Onose, N. Kanazawa, J. H. Park, J. H. Han, Y. Matsui, N. Nagaosa, and Y. Tokura, Real-space observation of a two-dimensional skyrmion crystal, *Nature* **465**, 901 (2010).

[118] D. Lu, D. J. Baek, S. S. Hong, L. F. Kourkoutis, Y. Hikita, and H. Y. Hwang, Synthesis of freestanding single-crystal perovskite films and heterostructures by etching of sacrificial water-soluble layers, *Nature Materials* **15**, 1255 (2016).

[119] M. Li, C. Tang, T. R. Paudel, D. Song, W. Lü, K. Han, Z. Huang, S. Zeng, X. Renshaw Wang, P. Yang, Ariando, J. Chen, T. Venkatesan, E. Y. Tsymbal, C. Li, and S. J. Pennycook, Controlling the Magnetic Properties of LaMnO$_3$/SrTiO$_3$ Heterostructures by Stoichiometry and Electronic Reconstruction: Atomic-Scale Evidence, *Advanced Materials* **31**, 1901386 (2019).

[120] M. Varela, M. P. Oxley, W. Luo, J. Tao, M. Watanabe, A. R. Lupini, S. T. Pantelides, and S. J. Pennycook, Atomic-resolution imaging of oxidation states in manganites, *Physical Review B* **79**, 085117 (2009).

[121] W. Luo, A. Franceschetti, M. Varela, J. Tao, S. J. Pennycook, and S. T.

Pantelides, Orbital-occupancy versus charge ordering and the strength of electron correlations in electron-doped CaMnO$_3$, *Physical Review Letters* **99**, 036402 (2007).

[122] P. G. Radaelli, D. E. Cox, M. Marezio, and S.-W. Cheong, Charge, orbital, and magnetic ordering in La$_{0.5}$Ca$_{0.5}$MnO$_3$, *Physical Review B* **55**, 3015 (1997).

[123] M. R. Ibarra, P. A. Algarabel, C. Marquina, J. Blasco, and J. García, Large magnetovolume effect in yttrium doped La-Ca-Mn-O perovskite, *Physical Review Letters* **75**, 3541 (1995).

[124] C. Ritter, M. R. Ibarra, J. M. De Teresa, P. A. Algarabel, C. Marquina, J. Blasco, J. García, S. Oseroff, and S.-W. Cheong, Influence of oxygen content on the structural, magnetotransport, and magnetic properties of LaMnO$_{3+x}$, *Physical Review B* **56**, 8902 (1997).

[125] J. De Teresa, M. Ibarra, J. Blasco, J. García, C. Marquina, P. Algarabel, Z. Arnold, and K. Kamenev, Spontaneous behavior and magnetic field and pressure effects on La$_{2/3}$Ca$_{1/3}$MnO$_3$ perovskite, *Physical Review B* **54**, 1187 (1996).

[126] M. Saghayezhian, S. Kouser, Z. Wang, H. Guo, R. Jin, J. Zhang, Y. Zhu, S. T. Pantelides, and E. W. Plummer, Atomic-scale determination of spontaneous magnetic reversal in oxide heterostructures, *Proceedings of the National Academy of Sciences of the United States of America* **116**, 10309 (2019).

[127] E. J. Moon, Q. He, S. Ghosh, B. J. Kirby, S. T. Pantelides, A. Y. Borisevich, and S. J. May, *Physical Review Letters* **119**, 197204 (2017).

[128] A. Vailionis, H. Boschker, W. Siemons, E. P. Houwman, D. H. A. Blank, G. Rijnders, and G. Koster, Structural δ doping to control local magnetization in isovalent oxide heterostructures, *Physical Review B* **83**, 064101 (2011).

[129] K. Zhao, J. Taylor, L. Habor, J. Zhang, E.W. Plummer, and M. Saghayezhian, Probing the interfacial symmetry using rotational second-harmonic generation in oxide heterostructures, *The Journal of Physical Chemistry C* **123**, 23000 (2019).

[130] A. Rockett, *The Materials Science of Semiconductors* (Springer US, Boston,

MA, 2008), pp. 289–356.

[131] H. Hilgenkamp and J. Mannhart, Grain boundaries in high-Tc superconductors, *Reviews of Modern Physics* **74**, 485 (2002).

[132] S. V Kalinin and N. A. Spaldin, Functional ion defects in transition metal oxides, *Science* **341**, 858 (2013).

[133] D. B. Holt and B. G. Yaobi, *Extended Defects in Semiconductors: Electronic Properties, Device Effects and Structures.* (Cambridge University Press, New York, USA, 2014).

[134] Richard J.D. Tilley, *Defects in Solids* (John Wiley & Sons, Inc., Hoboken, New Jersey, 2008).

[135] M. A. Zurbuchen, W. Tian, X. Q. Pan, D. Fong, S. K. Streiffer, M. E. Hawley, J. Lettieri, Y. Jia, G. Asayama, S. J. Fulk, D. J. Comstock, S. Knapp, A. H. Carim, and D. G. Schlom, Morphology, structure, and nucleation of out-of-phase boundaries (OPBs) in epitaxial films of layered oxides, *Journal of Materials Research* **22**, 1439 (2007).

[136] S. Celotto, W. Eerenstein, and T. Hibma, Characterization of anti-phase boundaries in epitaxial magnetite films, *The European Physical Journal B - Condensed Matter and Complex Systems* **36**, 271 (2003).

[137] D. Gilks, L. Lari, Z. Cai, O. Cespedes, A. Gerber, S. Thompson, K. Ziemer, and V. K. Lazarov, Magnetism and magnetotransport in symmetry matched spinels: $Fe_3O_4/MgAl_2O_4$, *Journal of Applied Physics* **113**, 17B107 (2013).

[138] Z. Wang, H. Guo, S. Shao, M. Saghayezhian, J. Li, R. Fittipaldi, A. Vecchione, P. Siwakoti, Y. Zhu, J. Zhang, and E. W. Plummer, Designing antiphase boundaries by atomic control of heterointerfaces, *Proceedings of the National Academy of Sciences of the United States of America* **115**, 9485 (2018).

[139] J. S. Jeong, M. Topsakal, P. Xu, B. Jalan, R. M. Wentzcovitch, and K. A. Mkhoyan, A new line defect in $NdTiO_3$ perovskite, *Nano Letters* **16**, 6816 (2016).

[140] N. B. Hannay, editor, *Treatise on Solid State Chemistry, Volume 3, Crystalline and Noncrystalline Solids* (Plenum Press, New York - London, 1976).

[141] H. S. Walter Greiner, Ludwig Neise, *Thermodynamics and Statistical Mechanics* (Springer-Verlag, 1995).

[142] Y.-M. Kim, J. He, M. D. Biegalski, H. Ambaye, V. Lauter, H. M. Christen, S. T. Pantelides, S. J. Pennycook, S. V Kalinin, and A. Y. Borisevich, Probing oxygen vacancy concentration and homogeneity in solid-oxide fuel-cell cathode materials on the subunit-cell level, *Nature Materials* **11**, 888 (2012).

[143] Y. Zhu, J. M. Zuo, A. R. Moodenbaugh, and M. Suenaga, Grain-boundary constraint and oxygen deficiency in $YBa_2Cu_3O_{7-\delta}$: Application of the coincidence site lattice model to a non-cubic system, *Philosophical Magazine A* **70**, 969 (1994).

[144] D. A. Muller, N. Nakagawa, A. Ohtomo, J. L. Grazul, and H. Y. Hwang, Atomic-scale imaging of nanoengineered oxygen vacancy profiles in $SrTiO_3$, *Nature* **430**, 657 (2004).

[145] O. L. Krivanek, T. C. Lovejoy, M. F. Murfitt, G. Skone, P. E. Batson, and N. Dellby, Towards sub-10 meV energy resolution STEM-EELS, *Journal of Physics: Conference Series* **522**, 12023 (2014).

[146] S. Wang, K. March, F. A. Ponce, and P. Rez, Identification of point defects using high-resolution electron energy loss spectroscopy, *Physical Review B* **99**, 115312 (2019).

[147] D. S. Su, H. W. Zandbergen, P. C. Tiemeijer, G. Kothleitner, M. Hävecker, C. Hébert, A. Knop-Gericke, B. H. Freitag, F. Hofer, and R. Schlögl, High resolution EELS using monochromator and high performance spectrometer: comparison of $V_2O_5$ ELNES with NEXAFS and band structure calculations, *Micron* **34**, 235 (2003).

[148] M. Bugnet, D. Rossouw, G. A. Botton, and T. Kolodiazhnyi, High-Resolution Near-Edge Structures in $EuTiO_3$, $SrTiO_3$ and $BaTiO_3$, *Microscopy and Microanalysis* **18**, 1460 (2012).

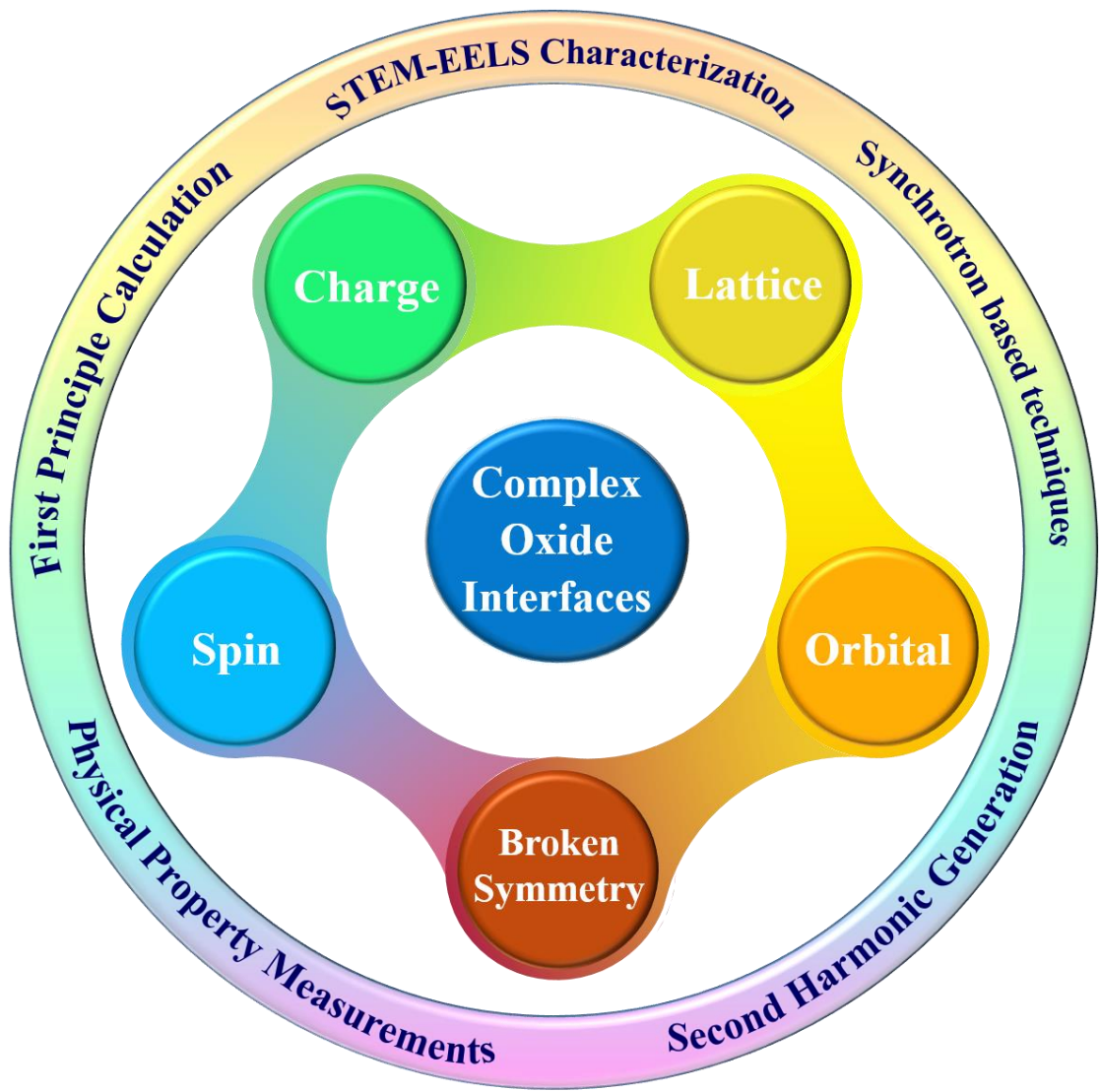

**Fig. 1 Schematics of ingredients in complex oxide interfaces.** Other than charge, spin, lattice and orbital degrees of freedom, broken symmetry plays the key role. The development of STEM-EELS techniques greatly advances our understanding of complex oxide interfaces.

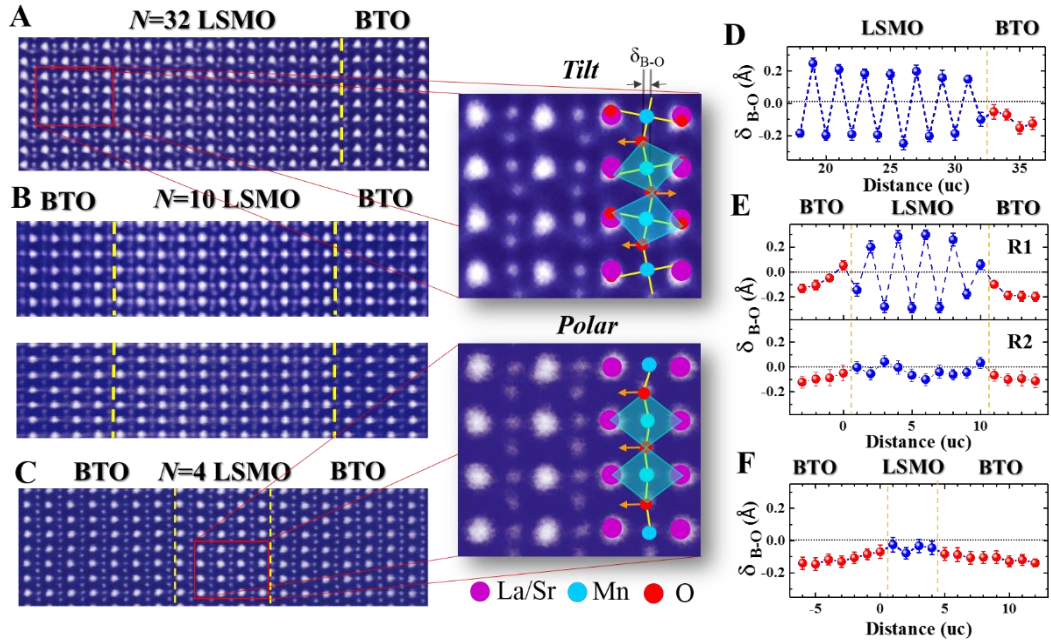

**Figure 2 Structural evolution of BTO(21)/LSMO(N) superlattices.** Intensity-inversed ABF-STEM image of BTO/LSMO superlattice with (A) $N = 32$, (B) $N = 10$ and (C) $N = 4$ u.c. taken along [110] direction. Enlarged images showing the $MnO_6$ octahedral AFD tilt in (A) and Mn-O polar displacements in (C) for LSMO film, with the atomic model superimposed. (D-F) Metal cation-oxygen displacements ($\delta_{B-O}$, B=Mn, Ti) as a function of distance measured from the ABF-STEM image. Adapted from [61].

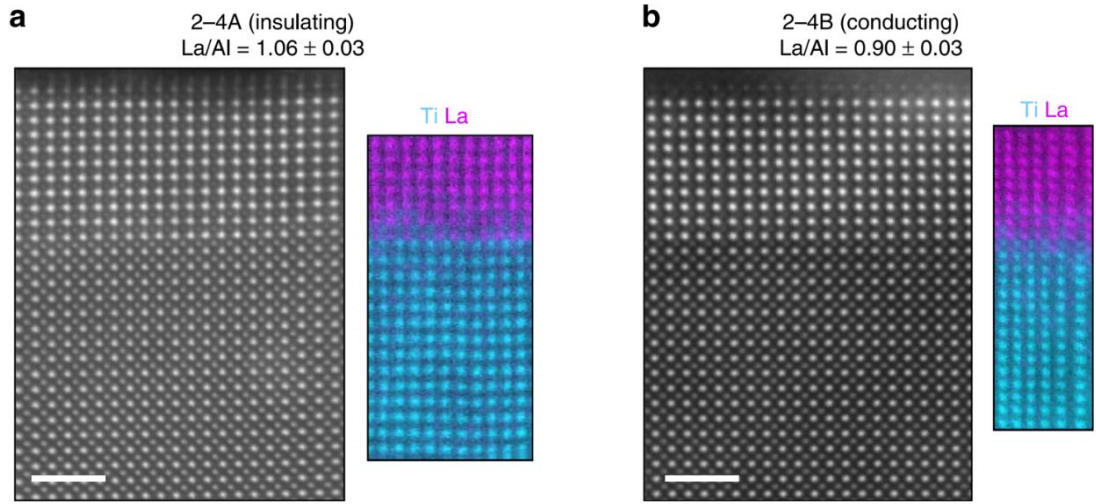

**Fig. 3 Interfacial La/Al concentration maps via HADDF STEM images and EELS spectroscopy.** (a) An insulating sample with La/Al ratio of 1.06 ± 0.03; (b) A conducting 2DEG sample with La/Al ratio of 0.90 ± 0.03. The EELS spectroscopic images map the concentration of lanthanum in magenta and titanium in turquoise; both samples show a small amount of interdiffusion at the interface. Reprinted by permission from [79]. Copyright © 2013 Nature Publishing Group.

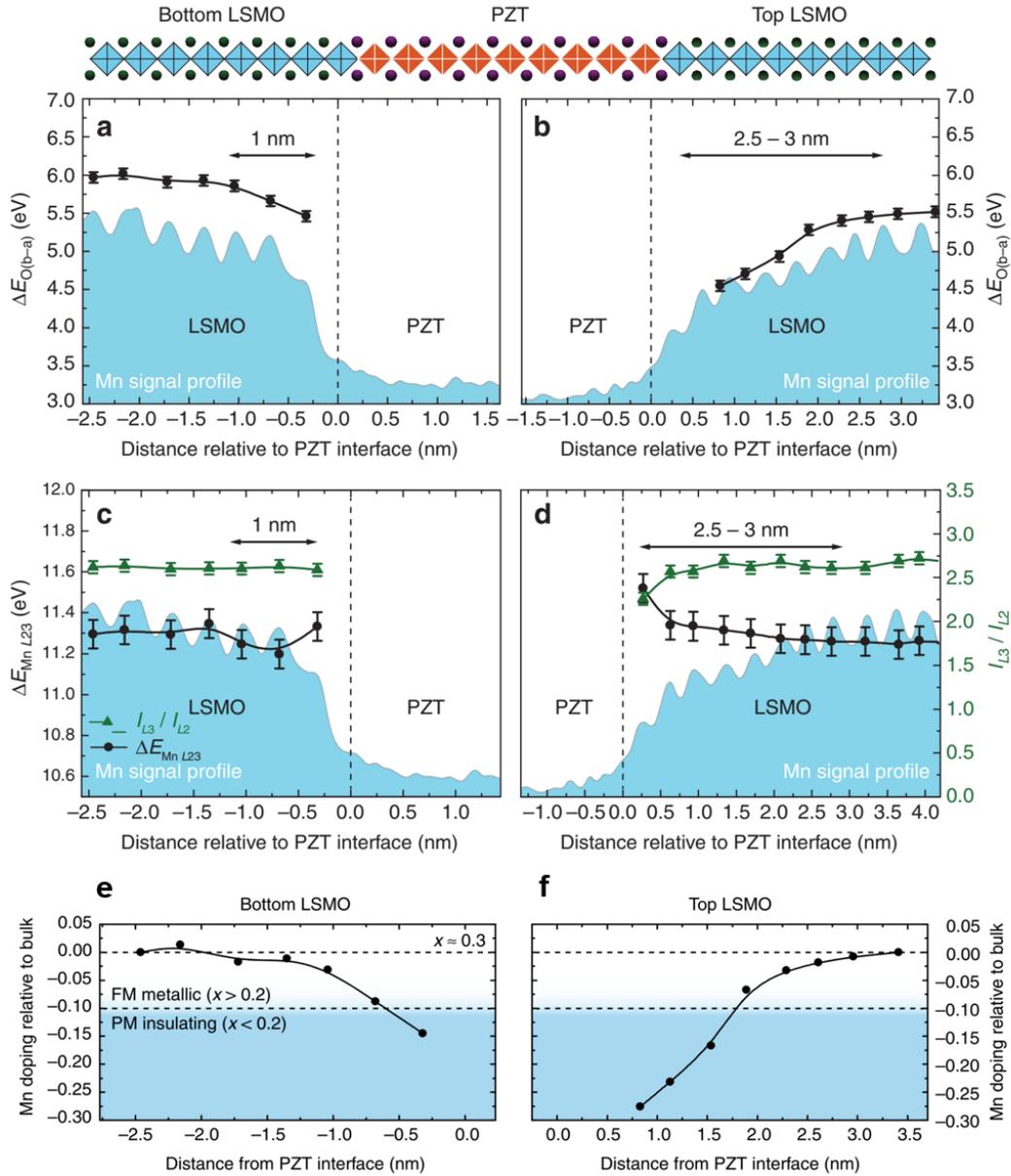

**Fig. 4 EELS measurements and Mn valence across LSMO/PZT/LSMO heterostructure.** (a,b) The O K pre- to main-peak separation in the vicinity of the PZT interface for the bottom and top LSMO layers, respectively. (c,d) The difference in Mn $L_{2,3}$ edge peak position (black circles) and $L_3/L_2$ peak intensity ratio (green triangles) in the vicinity of the PZT interface for the bottom and top LSMO layers, respectively. (e,f) Map of local Mn doping relative to bulk LSMO as a function of position normal to the LSMO/PZT interface for the bottom and top LSMO layers. Reproduced by permission from [75]. Copyright © 2015 Nature Publishing Group.

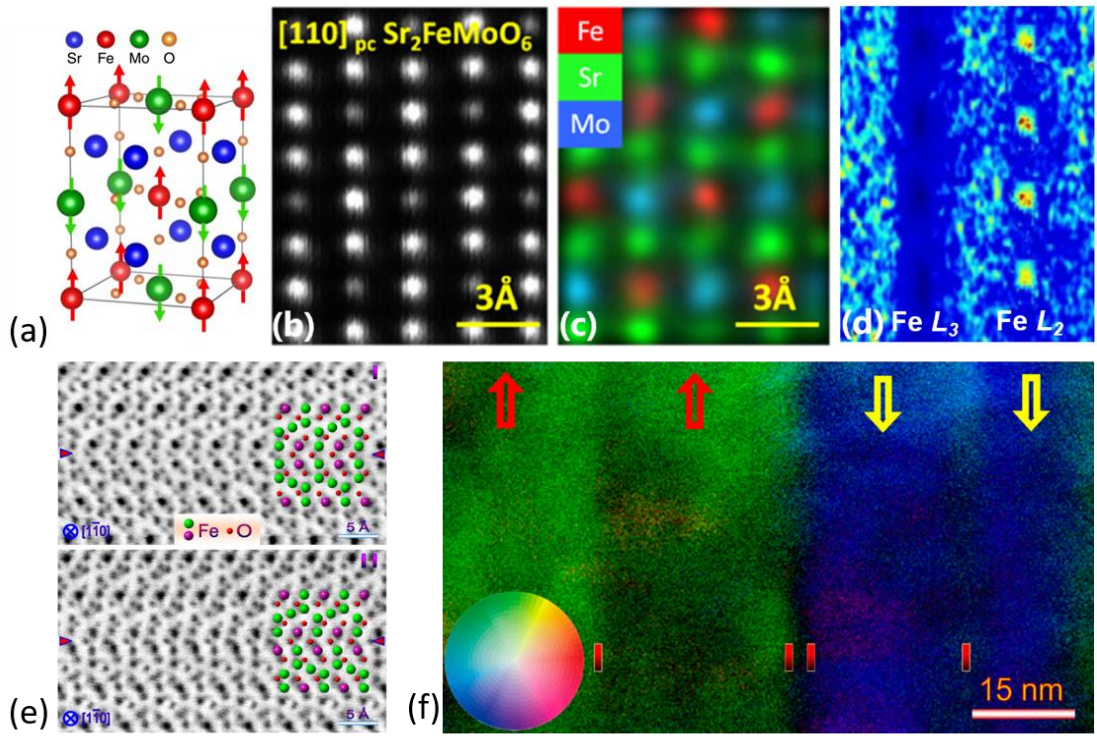

**Fig. 5 Identifying magnetic structure by DPC-STEM and EMCD.** (a) Crystal and magnetic structure of $Sr_2FeMoO_6$. (b) HAADF image and (c) EELS elemental maps along the [110]pc direction. (d) Atomic resolved EMCD of Fe, providing orbital and spin magnetic moments information. (e) ABF images $Fe_3O_4$ along [1-10] direction, showing Type I and II TBs. (f) Reconstructed in-plane magnetization vector maps of the two TBs obtained by DPC-STEM. (a-d): Reproduced by permission from [116]. Copyright © 2018 Nature Publishing Group. (e-f): Reprinted with permission from [113]. Copyright (2018) American Chemical Society.

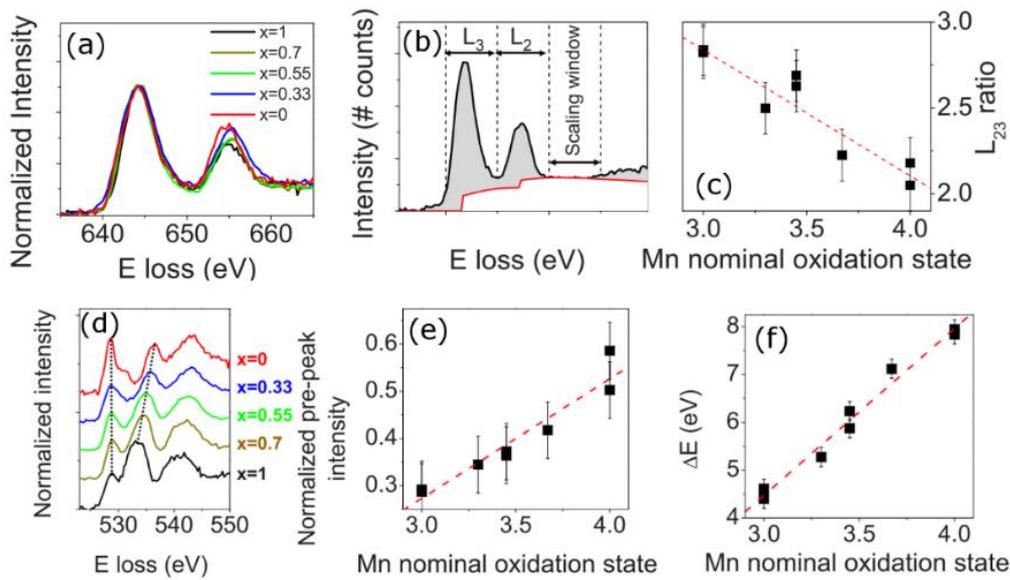

**Fig. 6 EELS and valence state in LCMO** (a) Mn L-edge in bulk $La_{1-x}Ca_xMnO_3$. To better see the intensity of $L_2$ and $L_3$ changing, the spectra are shifted and normalized. (b) Schematic spectra showing $L_3$ and $L_2$ edges with Hartree-Slater cross section step function (red). (c) $L_{23}$ ratio changes plotted against nominal oxidation state (from simple charge counting). (d) O K edges in bulk $La_{1-x}Ca_xMnO_3$ and the spectra are shifted so the pre-peaks are aligned. (e) Normalized pre-peak intensity vs nominal oxidation state. (f) Relative energy separation of pre-peak and main peak in O K edge as a function of Mn nominal oxidation state [120]. Reprinted with permission from [120]. Copyright (2009) by the American Physical Society.

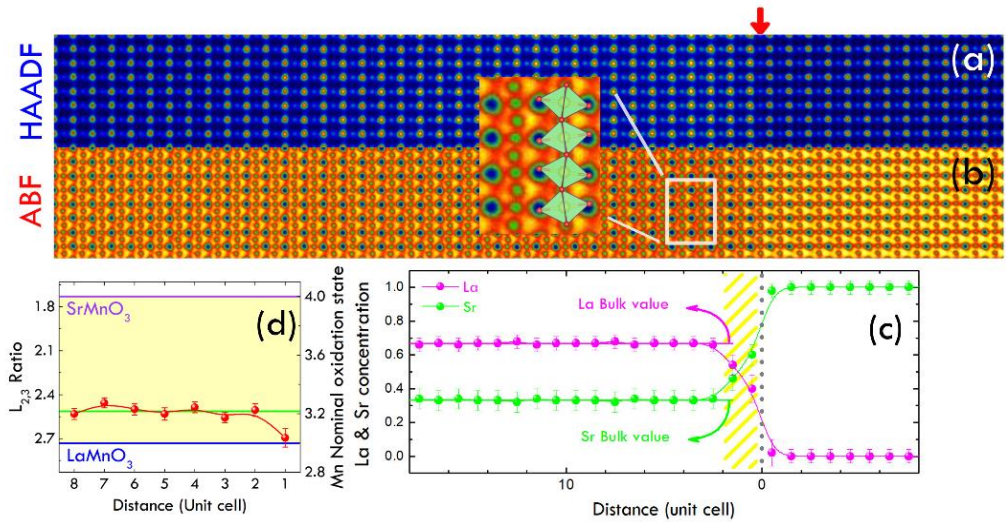

**Fig. 7 STEM characterization of LSMO/STO interface.** (a-b) HAADF and ABF STEM image of LSMO/STO along the [1-10] direction. The red arrow marks the interface. Inset shows the octahedral tilt away from the interface. (c) Layer-by-layer La and Sr concentration as function of distance from the interface. La and Sr reach their stoichiometric concentration, i.e. $La_{0.67}Sr_{0.33}$ after the third atomic column from the interface. (d) $L_{2,3}$ ratio and Mn nominal oxidation state as a function of distance from the interface [126]. Adapted from [126]

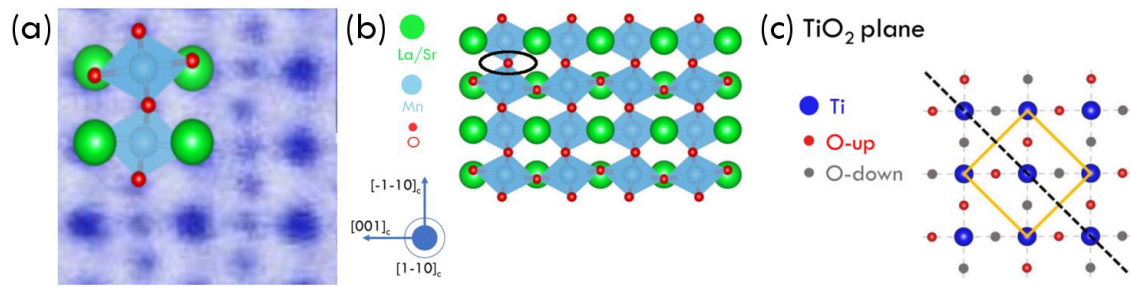

**Fig. 8 Symmetry at the LSMO/STO interface** (a) ABF-STEM image of LSMO/STO (001). The oxygen atoms show a zig-zag pattern. (b) The schematic of LSMO thin film with $a^-a^-c^0$ rotation pattern replicating the same zig-zag pattern. (c) TiO$_2$ plane with a reconstructed 2-dimensional lattice where only a mirror symmetry is present. Reprinted with permission from [129]. Copyright (2019) American Chemical Society.